\documentstyle[12pt,aps,epsfig]{revtex}
\def\Journal#1#2#3#4{{#1} {\bf #2}, #3 (#4)}
\def\NP{{\em Nucl. Phys.}}
\def\PL{{\em Phys. Lett.}}
\def\PR{{\em Phys. Rep.}}
\def\PRL{\em Phys. Rev. Lett.}
\def\PRD{{\em Phys. Rev.} D}
\def\PRC{{\em Phys. Rev.} C}
\def\ZP{{\em Z. Phys.}}

\newcommand{\be}{\begin{equation}}
\newcommand{\ee}{\end{equation}}
\newcommand{\bea}{\begin{eqnarray*}}
\newcommand{\eea}{\end{eqnarray*}}

\begin{document}
\title{Direct photons in Pb+Pb at CERN-SPS
from microscopic transport theory}
\author{A.\ Dumitru, M.\ Bleicher, S.A.\ Bass, C.\ Spieles,
L.\ Neise, H.\ St\"ocker, W.\ Greiner}
\address{Institut f\"ur Theoretische Physik der J.W. Goethe-Universit\"at\\
Postfach 111932, D-60054 Frankfurt a.M., Germany}
\date{\today}
\maketitle

\begin{abstract}
Direct photon production in central Pb+Pb collisions at CERN-SPS energy
is calculated within the relativistic microscopic transport model
{\small UrQMD}, and within distinctly different versions of relativistic
hydrodynamics. We find that in {\small UrQMD} the local momentum distributions
of the secondaries are strongly elongated along the beam axis initially.
Therefore, the pre-equilibrium contribution dominates
the photon spectrum at transverse momenta above $\approx 1.5$~GeV. The
hydrodynamics prediction of a strong correlation between the temperature and
radial expansion velocities on the one hand and the slope of the transverse
momentum distribution of direct photons on the other hand thus is not
recovered in {\small UrQMD}. The rapidity distribution of
direct photons in {\small UrQMD} reveals that the initial conditions for the
longitudinal expansion of the photon source (the meson ``fluid'') resemble
rather boostinvariance than Landau-like flow.
\end{abstract}
\newpage

The radiation of real and virtual photons has frequently been proposed
as a diagnostic tool for the hot and dense matter
created in (ultra-) relativistic heavy-ion collisions \cite{ph,KLS}.
The mean free path of photons with high transverse momentum exceeds the
expected source sizes by one order of magnitude \cite{KLS,Thoma}. Hard photons
thus can give insight into the early stage of these reactions.

Transverse momentum dependent upper limits for direct photon production in
central S+Au collisions
at $200A$~GeV have been published by the WA80 collaboration \cite{WA80}.
These data initiated theoretical studies \cite{Srv,LBK,CRS,Dum95a} which
showed that direct photon production is strongly overestimated if the
photon source is thermalized (having an initial energy density of
$2-5~{\rm GeV/fm^3}$) and if one assumes that it is composed of light mesons
only (say $\pi$, $\eta$, $\rho$, $\omega$). Due to the low specific entropy
of these particles a reasonable final-state multiplicity of secondaries
and a maximum temperature that is consistent with the WA80 data
($T_{max}\le300$~MeV) can not be achieved simultaneously (at least if the
expansion is approximately isentropic).

Most of the calculations quoted (except those of ref.\ \cite{LBK}), however,
assumed that the photon source is in local thermal equilibrium and that
ideal hydrodynamics can be applied to determine the space-time evolution
of the temperature. Here, we perform a calculation within the microscopic
transport model {\small UrQMD}, which includes the pre-equilibrium
contributions to direct
photon production. This is of particular relevance in view of the fact that
local thermalization (i.e.\ locally isotropic momentum distributions) in
(ultra-) relativistic heavy-ion collisions is probably not achieved within
the first few fm/c \cite{Sorge,Mattiello} where the high-$k_T$ photons
are produced.
Since neither thermal nor chemical equilibrium is assumed,
the effects of finite viscosity (i.e.\ finite mean free paths) \cite{visc}
and inhomogeneous or fluctuating energy
density distributions \cite{Zhang} are included.
Comparisons of the transverse momentum and
rapidity distributions of direct photons with those calculated within
various fluid-dynamical models are made.

The calculations presented here are based on the {\small UrQMD} model
\cite{urqmd}, which spans the
entire presently available range of energies from SIS to SPS.
It's collision term contains 50 different baryon species
(including nucleon, delta and hyperon resonances with masses up to $2.2$~GeV)
and 25 different meson species (including strange meson resonances), which
are supplemented by their corresponding antiparticles
and all isospin-projected states.

\begin{figure}
\centerline{\hbox{\epsfig{figure=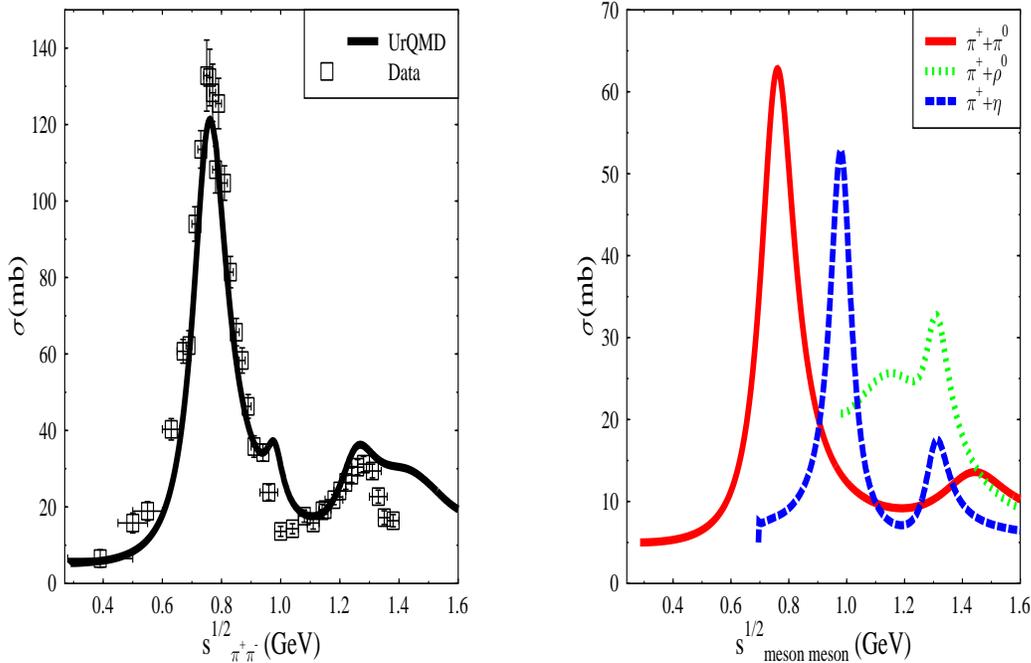,height=11cm,width=15cm}}}
\caption{Total cross sections of $\pi^+\pi^-$ and various other meson-meson
reactions
as calculated within the {\small UrQMD} model, data from ref.\
\protect\cite{proto}.}
\label{sigmm}
\vspace*{.5cm}
\end{figure}
The model is based on the covariant propagation of all hadrons on classical
trajectories, excitation of resonances and strings and their
subsequent decay resp.\ fragmentation. {\small UrQMD} accounts for
secondary interactions: the annihilation
of produced mesons on baryons can lead to the formation of $s$-channel
resonances or strings. Free cross-sections for hadron-hadron
scattering in the hot and dense nuclear matter are employed. Comparisons of
{\small UrQMD} calculations to experimental hadron yields from 
SIS to SPS are documented elsewhere \cite{urqmd}.

We briefly discuss the two most important reactions for the creation
of direct photons. Below $2$~GeV center of mass energy
intermediate resonance states are excited. The total cross section for
these reactions are given by
\begin{eqnarray}
\sigma^{M_1M_2}_{\rm tot}(\sqrt{s}) &=& \sum\limits_{R=\rho,f_2,\dots}
       \langle j_{M_1}, m_{M_1}, j_{M_2}, m_{M_2} \| J_R, M_R \rangle \,
        \frac{2 I_R +1}{(2 I_{M_1} +1) (2 I_{M_2} +1 )}  \nonumber\\
&&\times        \frac{\pi}{p^2_{CM}}\,
        \frac{\Gamma_{R \rightarrow M_1M_2} \Gamma_{\rm tot}}
             {(M_R - \sqrt{s})^2 + \frac{\Gamma_{\rm tot}^2}{4}}\quad,
\end{eqnarray}
which depends on the total decay width $\Gamma_{\rm tot}$, the partial decay
width $\Gamma_{R \rightarrow M_1M_2}$ and on the c.m. energy $\sqrt{s}$.
At higher energies these processes become less important. One then enters the
region of t-channel scattering of the hadrons. 

{\small UrQMD} predicts a rich and non-trivial resonance structure in the
meson-meson scattering cross section.
In fig.\ \ref{sigmm} (left) the total $\pi^+\pi^-$ cross section as calculated
within {\small UrQMD} is compared to experimental data \cite{proto}.
Below $1$~GeV the spectrum is dominated by the
$\rho$ resonance, while at higher energy the $f_2(1270)$ gets important.

\begin{figure}
\vspace*{-2cm}
\centerline{\hbox{\epsfig{figure=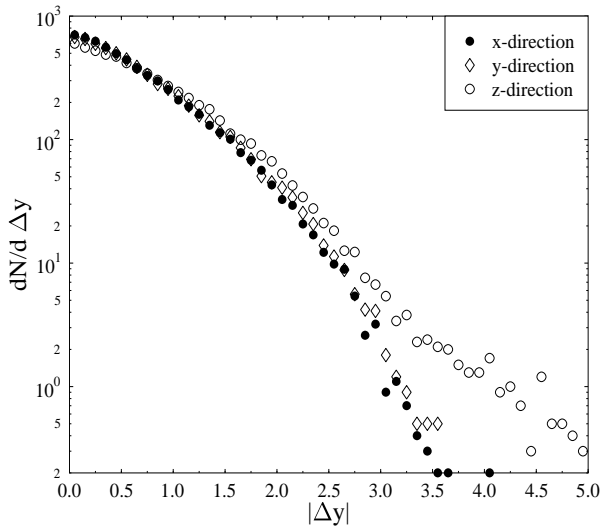,height=11cm,width=15cm}}}
\vspace*{-1cm}
\caption{Number of $\pi\pi$ collisions as a function of the rapidity
difference in the three spacial directions (Pb+Pb, $E_{Lab}=160A$~GeV,
$b=0$~fm).}
\label{dNddy}
\vspace*{.5cm}
\end{figure}
Figure \ref{dNddy} shows the distribution of $\pi\pi$ collisions
with a given relative rapidity in a given spacial direction (upper indices
refer to the particle number):
\be \label{DY_3D}
\Delta y_{x,y,z} = \frac{1}{2}
\ln \frac{E^1+p^1_{x,y,z}}{E^1-p^1_{x,y,z}} -
\frac{1}{2}\ln \frac{E^2+p^2_{x,y,z}}{E^2-p^2_{x,y,z}}
\quad.\ee
One observes that most collisions occur at $|\Delta y|<3$. In this range
the collision-spectrum is nearly isotropic, corresponding to a temperature of
$T=180\pm30$~MeV. In the longitudinal direction, however, there is an
additional component above $|\Delta y_z|=3$. This is due to pions created in
the fragmentation of a color flux tube close to projectile and target
rapidities. In p($200A$~GeV)+p collisions $\approx15\%$ of all pions have
$|y^\pi_{CM}|>2$ \cite{pi15}. Note that not all of these collisions are
forbidden by the finite formation time of secondaries since the
fragmentation region mesons contain one of the initial valence quarks
(i.e.\ the sources of the color field). Thus, even within the formation time,
these mesons may interact with half of the free meson cross section in the
{\small UrQMD} model.
Obviously, these meson-meson collisions with large longitudinal
rapidity difference are not taken into account in hydrodynamical
calculations. However, as will be discussed below, it is just their
contribution which dominates the direct photon spectrum above
$k_T\approx1.5$~GeV.

Let us now get to the reference equilibrium calculations of direct photon
production within hydrodynamics. Here we assume either a three-dimensional
expansion with cylindrical symmetry and longitudinal boost invariance \cite{Bj}
or a one-dimensional expansion with Landau initial conditions
(i.e.\ $v=0$ at time $t=0$). In both cases the dynamical creation
of the fluid of secondaries in space-time is ignored and only the expansion
stage is treated. 

In the three-fluid model fluids one and two correspond to projectile and
target, while fluid three represents the newly produced particles around
midrapidity. Fluids one and two are coupled via source terms leading to energy
and momentum exchange. These interactions are due to binary collisions of the
nucleons in the respective fluids and are derived from nucleon-nucleon data.
A detailed discussion of this model as well as results (pion rapidity
and transverse momentum spectra at CERN-SPS, baryon stopping and directed
flow at AGS and SPS, the width of the compression shock waves etc.)
can be found in refs.\ \cite{Dum95a,trento,Brachi}.

In the hydrodynamical calculations the fluid of produced particles is assumed
to be the hottest, and thus the dominant source of high-$k_T$ photons.
It's equation of state is that of 
an ideal gas of massive $\pi$, $\eta$, $\rho$, and $\omega$ mesons below
$T_C=160$~MeV. For $T>T_C$ we assume an ideal QGP (massless, noninteracting
$u$, $d$ quarks and gluons) described within the MIT bag-model \cite{MIT}.
The bag constant is chosen such that the pressures of the two phases match
at $T=T_C$ ($B^{1/4}=235$~MeV) thus leading to a first order phase transition.

The number of emitted direct photons in each of the three phases
is parametrized as \cite{KLS}
\be \label{rate}
E\frac{dN^\gamma}{d^4x\,d^3k} = \frac{5\alpha \alpha_S}{18\pi^2}
T^2 e^{-E/T} \ln \left( \frac{2.912E}{g^2 T} +1 \right)   \quad.
\ee
$E$ is the photon energy in the local rest frame. In our calculations
we fix $\alpha_S=g^2/4\pi=0.4$. Eq.\ (\ref{rate}) accounts for
pion annihilation ($\pi\pi\rightarrow \rho\gamma$), and Compton-like
scattering ($\pi\rho\rightarrow \pi\gamma$, $\pi\eta\rightarrow \pi\gamma$)
off a $\rho$ or $\eta$ meson (in lowest order perturbation theory). In the
QGP phase, alternatively, quark-antiquark annihilation ($q\overline{q}
\rightarrow g\gamma$) and Compton-like scattering of a quark or antiquark
off a gluon ($q,\overline{q}+g\rightarrow q,\overline{q}+\gamma$) are
considered. Contributions from
the $A_1$ meson \cite{Song}, as well as the effect of hadronic
formfactors \cite{KLS}, are neglected since they are of the   
same magnitude as higher order corrections to eq.\ (\ref{rate}), which we
have also not taken into account. Also, the number of processes that
contribute to direct photon production enhances the rate only linearly, while
the temperature distribution enters exponentially. Our main interest here
is not the absolute photon yield but rather the effective temperature of the
direct photons, i.e.\ the inverse slope of their transverse momentum
spectrum.

The photon spectrum emitted in a heavy-ion collision
is obtained here by an (incoherent) integration over space-time:
\begin{equation} \label{spec}
\frac{dN^\gamma}{d^2k_T\,dy}=\int d^4x \, E\frac{dN^\gamma}{d^4x\,d^3k}
\quad.
\end{equation}
To allow for a comparison with the hydrodynamical calculations we have
considered the same processes also in the {\small UrQMD} model, namely
$\pi^\pm\pi^\mp \rightarrow \rho^0\gamma$,
$\pi^\pm\pi^0 \rightarrow \rho^\pm\gamma$,
$\pi^\pm\rho^0 \rightarrow \pi^\pm\gamma$,
$\pi^\pm\rho^\mp \rightarrow \pi^0\gamma$,
$\pi^0\rho^\pm \rightarrow \pi^\pm\gamma$,
$\pi^\pm\eta \rightarrow \pi^\pm\gamma$.
Of course, in {\small UrQMD} these processes are considered explicitly (we
employ the differential cross sections given in ref.\ \cite{KLS}) and are not
folded with thermal distribution functions, nor with a thermal
$\sqrt{s}$-distribution of meson-meson collisions. We find that the
$\pi\rho\rightarrow \pi\gamma$ and $\omega\rightarrow\pi\gamma$ processes are
dominant in the range $1$~GeV$\le k_T\le3$~GeV.
This is due to the fact that, in addition to the kinetic
energy, also the mass of the $\rho$ or $\omega$ can be converted into photon
energy. In contrast to the thermal rate (\ref{rate}), in the {\small UrQMD}
model $m_\rho$ and $m_\omega$ are smeared out according to a Breit-Wigner
distribution. For the
processes $\pi\pi\rightarrow \rho\gamma$ we have, however, assumed that
the $\rho$-meson in the final state is produced with the peak mass of
$770$~MeV. Since free current-quarks and gluons have not been included in the
present version of the {\small UrQMD} model, the
processes $q\overline{q}\rightarrow g\gamma$ etc.\ are not considered there.

\begin{figure}
\vspace*{-2cm}
\centerline{\hbox{\epsfig{figure=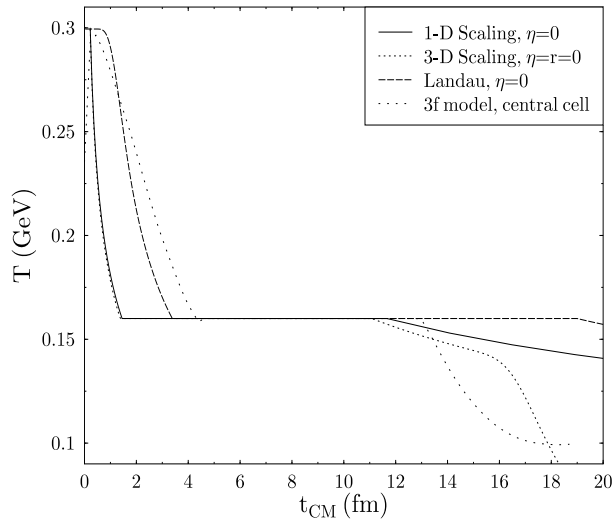,height=11cm,width=15cm}}}
\vspace*{-1cm}
\caption{Temperature in the central region as a function of time in various
hydrodynamical models, Pb+Pb-collisions ($b=0$~fm) at CERN-SPS.}
\label{cool}
\vspace*{.5cm}
\end{figure}
The temperature of the newly produced particles in the central region is
depicted in fig.\ \ref{cool} \cite{trento}. In case of a Bjorken
expansion we have assumed an initial temperature of $300$~MeV and
an initial time of $\tau_0=0.22$~fm/c, while for the Landau expansion
we take an initial longitudinal extension of $2L=2R_{\rm Pb}/\gamma_{CM}
\approx1.4$~fm. The maximum temperature in the
three-fluid model is determined by the coupling terms between fluids
one and two (and, of course, by the equation of state of fluid three).
The produced particles cool most swiftly in scaling hydrodynamics, while
in the case of a Landau expansion the temperature at midrapidity
is constant until the rarefaction waves reach the center.
In this latter case one finds the longest-living
mixed phase. This is due to the fact that we assumed purely one-dimensional
(longitudinal) expansion.

\begin{figure}
\vspace*{-2cm}
\centerline{\hbox{\epsfig{figure=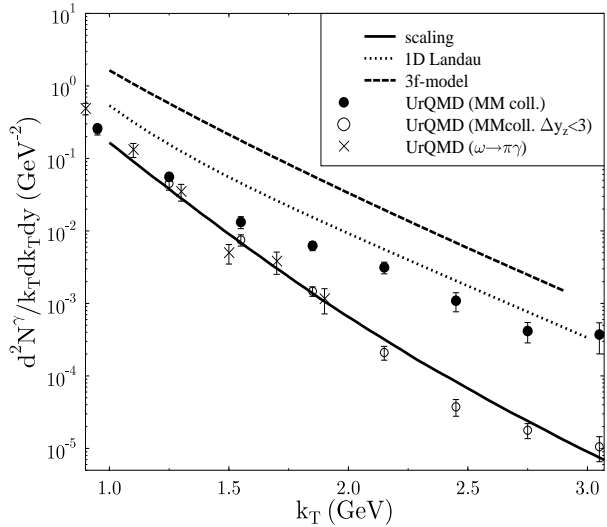,height=11cm,width=15cm}}}
\vspace*{-1cm}
\caption{Transverse momentum spectrum of direct photons at midrapidity
(Pb+Pb, $E_{Lab}=160A$~GeV, $b=0$~fm). The lines refer to hydrodynamical
calculations including a first order phase transition to a QGP at
$T_C=160$~MeV (three-fluid model (dashed),
scaling expansion (solid), and Landau expansion (dotted)).
The result of the {\small UrQMD} calculation with all
meson-meson collisions (full circles) is compared to the calculation with
only ``thermal'' meson-meson collisions (open circles). Crosses show the
photons from $\omega$-meson decays, which are as many as those
from ``thermal'' meson-meson collisions. Above $k_T\approx 1.5$~GeV the
pre-equilibrium contribution dominates.}
\label{phkt}
\vspace*{.5cm}
\end{figure}
The various transverse momentum distributions of direct photons,
cf.\ fig.\ \ref{phkt}, reflect the distinct cooling laws.
The faster the cooling of the photon source, the steeper the slope of
the photon spectrum. An exponential fit of the spectrum in the region
$2$~GeV$\le k_T\le3$~GeV yields $``T``=260$~MeV both for the three-fluid model
and for the Landau expansion, and $``T``=210$~MeV for the Bjorken expansion.

If we take into account only the ``thermal'' meson-meson collisions with
rapidity gap $|\Delta y_z|<3$, we find a similar photon spectrum in the
{\small UrQMD} model as in scaling hydrodynamics. The fact that the hadronic
mass
spectrum in {\small UrQMD} has no upper limit leads to a collision spectrum of
the light mesons that is equally ``cool'' as in the hydrodynamic
calculations, if there a first order phase transition at a critical temperature
around $160$~MeV is assumed (cf.\ also ref.\ \cite{CRS}). In UrQMD,
a significant part of the energy density can be stored into heavy meson-
and baryon resonances. This keeps the pressure and temperature low
\cite{Sorge,Bebie}. We have analyzed the temperature as a function of energy
density (at normal nuclear matter baryon density) in {\small UrQMD} in ref.\
\cite{urqmd}.

One also observes that the ``pre-equilibrium'' meson-meson collisions
with $|\Delta y_z|>3$ dominate the emission of direct photons with
large transverse momenta, $k_T>1.5$~GeV. Obviously, this contribution has
nothing to do with high meson transverse momenta (``temperatures'') or
fast collective radial expansion. It is thus not included in the
hydrodynamic calculations. Hence, the strong correlation between the
(Doppler-shifted)
temperature and the slope of the photon transverse momentum distribution
predicted by the hydrodynamical calculations (see also refs.\
\cite{Srv,CRS,Dum95a,Seibert}) is not seen in {\small UrQMD}.
Meson-baryon and baryon-baryon processes may further enhance the
pre-equilibrium contribution \cite{CSZP}. Below $k_T=1.5$ the
photons are dominantly produced by ``thermal'' meson-meson collisions and
$\omega\rightarrow\pi\gamma$ decays.

\begin{figure}
\vspace*{-2cm}
\centerline{\hbox{\epsfig{figure=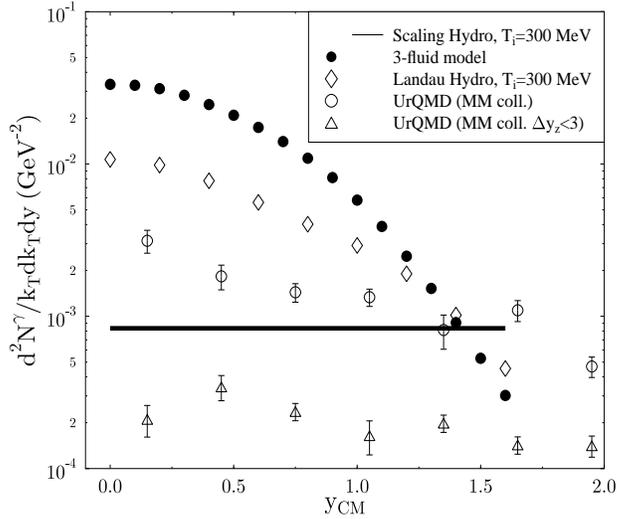,height=11cm,width=15cm}}}
\vspace*{-1cm}
\caption{Rapidity distribution of direct photons,
calculated within the various hydrodynamical models (for $k_T=2$~GeV)
and {\small UrQMD} (for $k_T=2.15$~GeV).}
\label{phy}
\vspace*{.5cm}
\end{figure}
The rapidity distribution of the secondaries (or, alternatively, of the
temperature) at early times is very different in the various hydrodynamical
models \cite{trento,Dum95b}. While $T(\eta)$ ($\eta$ denotes the fluid
rapidity) is strongly peaked around midrapidity in the three-fluid and
Landau models, scaling flow $v_z=z/t$ and energy-momentum conservation,
$\partial_\mu T^{\mu\nu}=0$, imply that the pressure $p(T,\mu_B)$ is
independent of the fluid rapidity, $\partial_\eta p=0$, and depends only on
proper time, $\tau=\sqrt{t^2-z^2}$. For net baryon-free matter
$\partial_\eta T=0$ follows. However, even in the first case matter is
accelerated by a (longitudinal)
rarefaction wave, leading to a broadening of the rapidity
distribution. Thus, in any case, at the late freeze-out stage the rapidity
distribution of secondaries (around midrapidity) will be more or less flat.
On the other hand, the rapidity distribution of direct photons with transverse
momenta much larger than the maximum temperature of the photon source (e.g.\
$k_T\approx2$~GeV at SPS) offers the opportunity to measure the rapidity
distribution of the hot photon source {\sl at early times}
\cite{Dum95b}. This is demonstrated in fig.\ \ref{phy}.
In the three-fluid model and in
the Landau-expansion the photon rapidity distribution is strongly
peaked around midrapidity. It is not proportional to the (squared) rapidity
distribution of the pions, which are emitted at the much later freeze-out
stage. The distribution obtained within {\small UrQMD} only resembles that for
a boostinvariant expansion, if only ``thermal''
meson-meson collisions are taken into account. The full calculation
predicts a maximum of the photon-$dN/dy$ at midrapidity. This is so because
the photon spectrum emitted
in elementary meson-meson reactions with large $\Delta y$ is peaked around
midrapidity. Nevertheless, the photon distribution is considerably flater
than in the three-fluid or Landau models, which show a concave curvature in
this logarithmic plot. This is a consequence of the longitudinal velocity
profile \cite{Dirk} induced by the (longitudinal)
rarefaction wave accelerating the fluid that is initially at rest.
The UrQMD calculation does not exhibit this characteristic feature.

In summary, we have calculated direct photon production in central
Pb+Pb collisions at CERN-SPS energy within a microscopic transport
model ({\small UrQMD}). Within this model, secondary hadron production is
described by
color flux tube fragmentation and resonance decay. This leads to non-thermal
momentum distributions, i.e.\ they are strongly elongated along the beam axis
initially. The contribution from this pre-equilibrium stage to direct photon
production is found to dominate at transverse momenta above $\approx1.5$~GeV.
Thus, within the {\small UrQMD} model, the slope of the photon transverse
momentum spectrum is not directly related to the Doppler-shifted ``transverse
temperature'' of the source, as it is in hydrodynamics.

If only ``thermal'' meson-meson collisions are taken into account, both
the transverse momentum and rapidity spectra of direct photons in
{\small UrQMD} resemble those calculated within boostinvariant hydrodynamics,
if for the latter one assumes an initial temperature $T_i=300$~MeV and initial
time $\tau_i=0.22$~fm. This is so although QGP formation is not assumed in
{\small UrQMD}. Hence, heavy resonances and color flux tubes allow for a rather
soft meson-meson collision spectrum even at high entropies.
\acknowledgements
This work was supported by Graduiertenkolleg Theoretische und Experimentelle
Schwer\-ionen\-physik, BMBF, DFG, and GSI.
M.\ Bleicher thanks the Josef Buchmann foundation for financial support.

\end{document}